\documentclass[12pt,twoside]{article}
\usepackage{fleqn,espcrc1,amsfonts}

\usepackage{graphicx}
\usepackage[figuresright]{rotating}

\newcommand{\re}{{\rm e}}
\newcommand{\ri}{{\rm i}}

\newcommand{\AmS}{{\protect\the\textfont2
  A\kern-.1667em\lower.5ex\hbox{M}\kern-.125emS}}

\title{Complex Scaling of the Faddeev Operator\thanks{This work is
supported in part by the Academia Sinica,
the National Science Council (ROC), and the Russian Foundation for
Basic Research.}
}

\author{E. A. Kolganova\address{Institute of Atomic and  Molecular
Sciences, Academia Sinica,\\ PO Box 23-166, Taipei, Taiwan 10764, Republic of
China}\address{Joint Institute for Nuclear Research, 141980 Dubna, Moscow region,
Russia},
A. K. Motovilov$^{\rm b}$
        and
        Y. K. Ho$^{\rm a}$}
\begin{document}

% typeset front matter
\maketitle

\begin{abstract}
The work is devoted to comparison of two different approaches to
calculation of three-body resonances on the basis of the Faddeev
differential equations. The first one is the well known complex
scaling approach. The second method is based on an immediate
calculation of the zeros of the scattering matrix continued to
the physical sheet.
\end{abstract}
\bigskip

\medskip

For years the complex scaling  \cite{a1,a2} is considered as one
of the most convenient methods for calculation of resonances.
This method gives a possibility to rotate the continuous
spectrum of an $N$-body Hamiltonian with analytic potentials.
The binding energies stay fixed in this rotation while the
resonances appear as additional complex eigenvalues of the
scaled Hamiltonian situated in the sectors of the unphysical
sheets which become accessible during the scaling
transformation. Surveys of the literature on the complex scaling
method and its various applications can be found in
%, e.\,g., in Refs.
\cite{Ho}. Here we only note that for a
rather wide class of the interaction potentials the
complex-scaling resonances are proved to be at the same time the
scattering-matrix resonances~\cite{Hagedorn} (i.\,e. they are
also the poles of the scattering matrix analytically continued
in corresponding unphysical sheets).

Various different methods, of course, are also used for
calculations of the resonances.  Among the methods developed to
calculate directly the scatering-matrix resonances we, first,
mention the approach based on the momentum space Faddeev
integral equations~\cite{MF} (see, e.\,g., Ref.~\cite{Orlov} and
references cited therein). In this approach one numerically
solves these equations continued into an unphysical sheet and,
thus, the three-body resonances arise as the poles of the
continued T-matrix.  Another approach to calculation of the
scattering-matrix resonances is based on the explicit
representations \cite{MotMN} for the T- and S-matrices,
continued into unphysical sheets, in terms of the physical
sheet. {}From these representations one infers that the
three-body resonances can be found as zeros of certain
truncations of the scattering matrix only taken in the physical
sheet. Such an approach can be employed even in the coordinate
representation \cite{MotMN,ourYaF1}.

To the best of our knowledge there are no published works
applying the scaling approach to the Faddeev equations. Thus,
the purpose of the present work is two-fold. On the one hand, we
make, for the first time, the complex scaling of the Faddeev
differential equations.  On the other hand we compare the
complex scaling method with the scattering-matrix
approach~\cite{MotMN,ourYaF1}.
%trying to undestand which of these
%methods is more efficient in the case of exponentially
%decreasing two-body potentials.
We do this making use of both
the approaches to examine a model system of three bosons
having the nucleon masses and the three-nucleon ($nnp$) system
itself.

%\section{Method}
We start with recalling that the three-body
Schr\"odinger operator $H_{3b}$ reads after the
scaling transorm as follows
\begin{eqnarray}
\label{SchrScal}
H_{3b}(\vartheta)=U(\vartheta)H_{3b}U(-\vartheta) =
-\re^{-2\vartheta}\Delta_X+
\sum_\alpha v_\alpha(\re^\vartheta |x_\alpha|),  \qquad \alpha=1,2,3.
\end{eqnarray}
where $\vartheta=\ri\theta$ with $\theta\in{\Bbb R}$ while
$x_\alpha,y_\alpha$ are the standard Jacobi variables,
$X\equiv(x_\alpha,y_\alpha)$. By $\Delta_X$ we understand the
6--dimensional Laplacian and by $v_\alpha$, the two-body
potentials.  The corresponding scaled Faddeev equations that we
solve have the following form:
\begin{eqnarray}
\label{FaddInhom}
[-\re^{-2\vartheta}\Delta_X+v_\alpha(\re^\vartheta |x_\alpha|)-z]
\Phi^{(\alpha)}(X)+
v_\alpha(\re^\vartheta |x_\alpha|)\sum_{\beta\neq\alpha}\Phi^{(\beta)}(X)
=f_\alpha(X),  \qquad \alpha=1,2,3.
\end{eqnarray}
Here $f=(f_1,f_2,f_3)$ is an arbitrary three-component vector
with components $f_\alpha$ belonging to the three-body Hilbert
space. The partial-wave version of the equations (\ref{FaddInhom}) for
a system of three identical bosons with  $L=0$ reads
\begin{eqnarray}
\label{FadPartCor}
{\re^{-2\ri\theta}} H_0^{(l)}\Phi_l(x,y) -z\,\Phi_l(x,y) +
V({\re^{\ri\theta}}x)\Psi_l(x,y)={{f_l(x,y)}},
\end{eqnarray}
Here, $H_0^{(l)}$ denotes the partial kinetic energy operator
and $\Psi_l$, the partial-wave component of the sum
$\Psi=\Phi^{(1)}+\Phi^{(2)}+\Phi^{(3)}$.  Explicit expression for
$H_0^{(l)}$ as well as a formula explicitly relating $\Phi_l$ to
$\Psi_l$ can be found, e.\,g., in~\cite{MF}.

For compactly supported inhomogeneus terms $f_l(x,y)$ the
partial-wave Faddeev component $\Phi_l(x,y)$ satisfies the
asymptotic condition
\begin{equation}
\label{HeBS}
    	\begin{array}{lll}
  \Phi_l(x,y) & = & \delta_{l0}\psi_d({\re^{\ri\theta}}x)\exp({\rm i}
   \sqrt{E_t-\epsilon_d}\,\,{\re^{\ri\theta}y}) \left[{\rm a}_0+
o\left(y^{-1/2}\right)\right] \\
        &+&
 \displaystyle\frac{\exp({\rm i}
 \sqrt{E_t}\,\,{\re^{\ri\theta}}\rho)}{\sqrt{\rho}}
 \left[A_l(y/x)+o\left(\rho^{-1/2}\right)\right] ,
\end{array}
\end{equation}
where $\psi_d(x)$ stands for the wave function of the two-body
subsystem while ${\rm a}_0$ and $A_l(y/x)$ the main asymptotical
coefficients effectively describing the contributions to
$\Phi_l$ from the elastic $(2\to2)$ and breakup $(2\to3)$
channels, respectively.

In the scaling method a resonance is looked for as the energy
$z$ which produces a pole to the function
\begin{equation}%
{\Phi(\theta,z)=\left<\left[H_F(\theta)-z\right]^{-1}
f,f\right>}\,
\end{equation}
where $H_F(\theta)$ is the non-selfadjoint operator resulting
from the complex-scaling transformation of the Faddeev operator.
This is just the operator constituted by the l.\,h.\,s. parts of
Eqs.  (\ref{FaddInhom}). The resonance energies should not, of
course, depend on the scaling parameter $\theta$ and on the
choice of the terms $f_l(x,y)$.

%\section{Results}

In the table we present our results obtained for a
complex-scaling resonance in the model three-body system which
consists of identical bosons having the nucleon mass. To
describe interacation between them we employ a Gauss-type
potential with a barrier term taken from Ref. \cite{ourYaF1} the
barrier value $V_b$ (see \cite{ourYaF1} for details) being
fixed, $V_b=1.5$. The figures in the table correspond to the
roots of the inverse function $[\Phi(\theta,z)]^{-1}$ calculated
on the basis of the finite-difference apporoximation of
Eqs.~(\ref{FadPartCor}\,--\,\ref{HeBS}) in polar coordinates.
Description of the finite-difference algorithm used can be found
in Ref. \cite{KMS-JPB}. In the present calculation we have taken
up to 400 knots in both hyperradius and hyperangle variables
while for the cut-off hyperradius we take 40\,fm.  One observes
from the table that the position of the resonance depends very
weakly on the scaling parameter $\theta$ which confirms a good
numerical quality of our results.

\begin{center}
\begin{tabular}{cccc}
\hline
$\theta$ & $z_{\rm res}$ (MeV) & $\theta$ & $z_{\rm res}$ (MeV) \\ \hline
0.25 & $-5.9525-0.4034\,{\rm i}$ & 0.50 & $-5.9526-0.4032\,{\rm i}$ \\ %\hline
0.30 & $-5.9526-0.4033\,{\rm i}$ & 0.60 & $-5.9526-0.4033\,{\rm i}$ \\ %\hline
0.40 & $-5.9526-0.4032\,{\rm i}$ & 0.70 & $-5.9526-0.4034\,{\rm i}$ \\ \hline
\end{tabular}
\end{center}

We compare the resonance values of the table to the resonance value
$z_{\rm res} =-5.952-0.403\,{\rm i}$\,MeV obtained for the same
three-boson system with exactly the same potentials but in the
completely different scattering-matrix approach of
Ref.~\cite{ourYaF1}. We see that, indeed, both the complex
scaling and the scattering matrix approaches give the same result.

As to the $nnp$ system where we employed the MT\,I--III~\cite{MT} model,
both these methods give no resonances on the two-body unphysical
sheet (see~\cite{ourYaF1}). Also we have found no resonances in
the part of the three-body sheet accessible via the scaling
method. Thus, at least in the framework of the MT\,I--III model we
can not confirm the experimatal result of Ref.~\cite{Alexan} in
which the point $-1.5\pm 0.3-\ri(0.3\pm0.15)$\,MeV was
interpreted as a resonance corresponding to an exited state of
the triton.

The triton virtual state can be only calculated within a
scattering-matrix method but not in the scaling approach.  Our
present improved scattering-matrix result for the triton virtual
state obtained within the method~\cite{MotMN,ourYaF1} is
$-2.690$\,MeV (i.\,e.  the virtual level lies
0.47\,MeV below the two-body threshold).  This result has been
obtained with the MT\,I-III potential on a grid having 1000
knots in both hyperradial and hyperradial variables and with the
value of cut-off hyperradius equal to 120\,fm.  Some values for the
energy of virtual state obtained by different authors can be
found in~\cite{Orlov}, and all of them are about 0.5MeV below
the two-body threshold.

\end{document}